# Measured and Modeled Outdoor Indoor Coverage at 28 GHz into High Thermal Efficiency Buildings

Dmitry Chizhik, Jinfeng Du, Reinaldo Valenzuela
*Nokia Bell Labs*, Murray Hill, NJ, U.S
dmitry.chizhik@nokia-bell-labs.com

Andrea Bedin, Martti Moisio
*Nokia Bell Labs*, Espoo, Finland

Rodolfo Feick
*Universidad Técnica Federico Santa Maria*, Chile

*Abstract*—28 GHz outdoor-indoor coverage into modern office buildings with high thermal efficiency windows is found to be severely limited due to 46 dB median penetration loss at normal incidence and additional 15 dB median oblique incidence loss. The study is based on measurements of path gain over 280 outdoor-indoor links, at ranges up to 100 m. A simple theoretical path gain model is extended to include building penetration through multiple sides of the building as well as a reflection from another building. The theoretical model accounts for the building orientation relative to the source, resulting in 4.9 dB RMSE relative to data, as compared to 5.7 dB RMSE from a linear fit and 14.7 dB RMSE for the 3GPP recommended model. Only coarse description of the buildings is required: building orientation and exterior wall composition, without any interior details. Coverage range for SNR>-8 dB from an outdoor base to a terminal just inside a high-efficiency building is under 35 m.

## I. Introduction

The commercial success of mm-wave bands for wireless communications [1] is critically dependent on achievable coverage range, largely determined by path gain. Accurate prediction of coverage is important for planning and deploying communication networks.

Path gain as a function of range is typically predicted using slope-intercept formulas, with parameters obtained from linear fit to measurements (path loss in dB w.r.t. distance in logarithmic scale) collected in similar environments [2]. Indoor terminals, for e.g in Fixed Wireless Access (FWA), suffer an additional building penetration loss [2]-[5], with strong dependence on building materials.

In this paper we present measurements and models of outdoor-indoor path loss collected at 28 GHz from rooftop mounted transmitters (representing base stations) to indoor terminals inside high thermal efficiency buildings. Exterior walls are concrete and low-E (low-emissivity) glass.

## II. Measurement Scenario

A 50° horn transmit antenna was placed on 4 locations on building rooftops (Tx1-Tx4 in Fig. 1), each emulating a base station. A cart-mounted receiver, representing an indoor terminal, was moved inside buildings, labeled 7A, 7B, 7C in Fig. 1. The building arrangement emulates an urban scenario, with buildings on both sides of a street. The receive antenna was a spinning 10° horn recording narrowband receive power, as a function of azimuth angle (equipment details in [6]). The receiver moved along an interior aisle, 6 m from the exterior wall, stopping every 1 m to collect measurements at ranges up to 100 m from the outdoor rooftop base station. Each building had 'North" and "South" aisles. Over 280 link measurements were made.

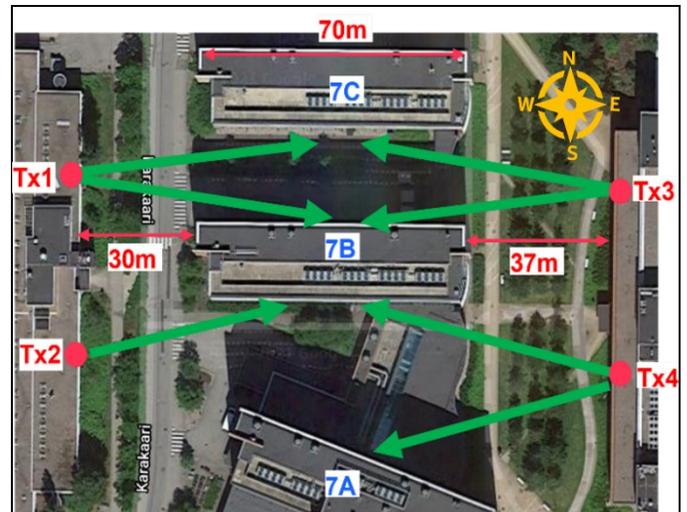

Fig 1. Outdoor-indoor measurement layout using 4 Tx locations on nearby rooftops. Indoor terminal moves at 1m step along interior aisles that are 6 m away from the building exterior windows.

Path gain is estimated from the azimuthal average received power and the beam pattern of the 50° transmit horn. The low-E glass windows of these buildings cover 30% of the wall area on north and south sides and entire east wall of 7B and 7C.

## III. Path Gain measurements and Models

Measured outdoor-indoor path gain values are plotted vs. range in Fig. 2 for two illustrative data subsets, with run-to-run spread exceeding 12 dB. The measurements are compared against an enhancement of the model [7] and a 3GPP model. Up to three paths (Fig. 3) are considered: direct illumination of 2 walls visible from base and a reflection from another building. Path gain for each is taken as (63) in [7].

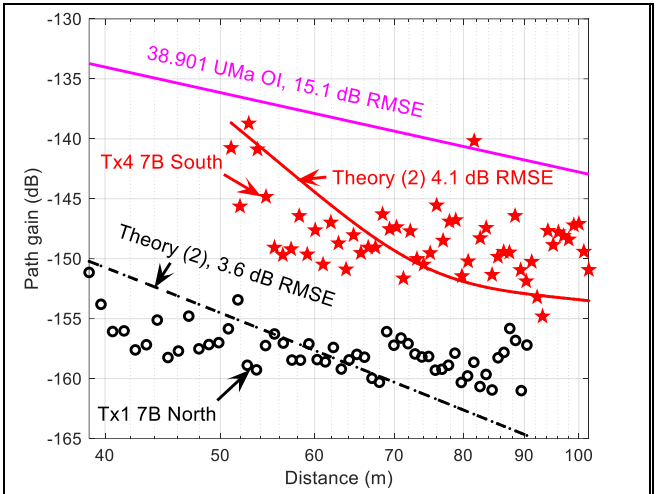

Fig. 2. Path gain data and models for Tx1-7B North and Tx4-7B South data.

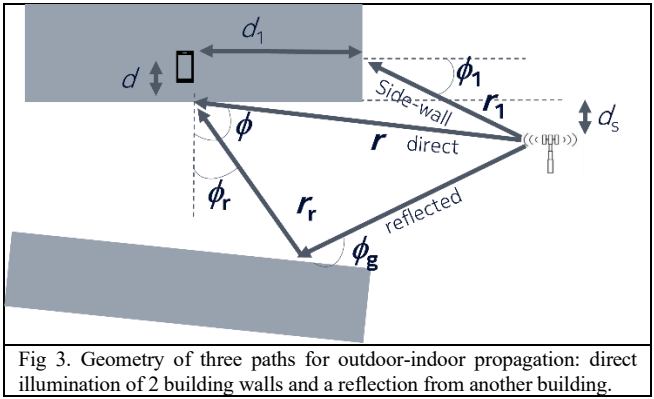

Fig 3. Geometry of three paths for outdoor-indoor propagation: direct illumination of 2 building walls and a reflection from another building.

$$P_{\text{O-I}} = \frac{\lambda^2 \cos^2 \phi \; T_{\text{eff}} e^{-\kappa_{\text{in}} d}}{8\pi^2 r^2} + \frac{\lambda^2 \cos^2 \phi_1 \; T_{\text{eff, side}} e^{-\kappa_{\text{in}} d_1}}{8\pi^2 (r_1 + d_1)^2}$$
$$+ \frac{\lambda^2 \cos^2 \phi_r \; T_{\text{eff}} \left|\Gamma_\perp (\phi_g)\right|^2 e^{-\kappa_{\text{in}} d}}{8\pi^2 r_r^2} \quad (1)$$

Where $\lambda$ is the wavelength, $r$ and $r_1$ are ranges from the base to points of exterior walls nearest the terminal and $r_r$ is the unfolded length of the reflected path, $\phi$, $\phi_1$, and $\phi_r$ the corresponding incident angles, $\cos^2 \phi = d_s^2/r^2$, with standoff distance $d_s$ and range $r$ (63) of [7]. $T_{\text{eff}} = 2.5 \times 10^{-5}$ (-46 dB) was estimated from separate short range window transmission measurements, with windows occupying 30% of wall area. For side-wall path ($r_1$, $\phi_1$), $T_{\text{eff, side}} = 1.5 \times 10^{-4}$ for floor-ceiling glass wall. The power reflection coefficient $\left|\Gamma_\perp\right|^2 = 0.3 + 0.7 e^{-(4/n_2)\phi_g}$ from the opposite building with 30% low-E glass wall, grazing angle $\phi_g$ and concrete refraction index $n_2 = \sqrt{5}$. The indoor absorption $\kappa_{\text{in}} = 0.12$ Nep/m (0.5 dB/m [2]). Only paths permitted by ray optics are included for each data set: the wall needs to be "visible" from the transmitter, reflection satisfying Snell's law.

For the 2 data subsets in Fig. 2, the path gain model (1) has RMSE under 4.1 dB, in contrast to 15.1 dB RMSE for 3GPP UMa O-I model [1]. The 3GPP UMa O-I high-loss model [1] was evaluated for 30% Low-E (IRR) glass wall (43 dB loss), and 3 dB indoor loss (0.5 dB/m [1], 6 m inside building).

Table 1 contains fit parameters and RMS errors for the 3GPP model and model (1) for each of 7 data subsets, and the combined data set. The model (1) achieves 4.9 dB RMSE overall in contrast to 14.7 dB RMSE for the 3GPP O-I model.

Table I. Outdoor-indoor model accuracy

| Data subset | Exponent $n$ | 1-m intercept (dB) | Fit RMSE (dB) | 3GPP RMSE (dB) | Theory RMSE (dB) |
|---|---|---|---|---|---|
| Tx 1 - 7B N | 1.00 | 139.7 | 1.5 | 19.5 | 3.6 |
| Tx 1 - 7C S | 2.24 | 114.6 | 2.4 | 16.7 | 4.6 |
| Tx 2 - 7B S | 1.15 | 137.9 | 1.6 | 20.4 | 4.5 |
| Tx 3 - 7B N | 2.05 | 116.1 | 1.9 | 14.6 | 6.8 |
| Tx 3 - 7C S | 3.74 | 83.8 | 2.5 | 14.0 | 5.4 |
| Tx 4 - 7B S | 1.39 | 122.2 | 2.7 | 8.7 | 4.1 |
| Tx 4 - 7A N | 0.75 | 132.3 | 4.3 | 8.4 | 5.6 |
| **Overall** | **0.68** | **140.1** | **5.7** | **14.7** | **4.9** |

Outdoor-indoor coverage range may be estimated using this path gain model and assuming transmit power of 30 dBm, base antenna gain 25 dBi, terminal gain 12 dBi, 100 MHz bandwidth and Noise figure NF=9 dB. It is found that SNR>-8 dB from an outdoor base to a terminal inside a high-efficiency building is under 35 m.

IV. CONCLUSIONS

Measured penetration loss into concrete buildings with low-E glass was found to be over 46 dB even at normal incidence. A simple path gain model was found to have an accuracy of 4.9 dB RMSE, in contrast to 14.7 RMSE from 3GPP model. Building orientation and exterior wall composition are sole inputs. Coverage range (for SNR>-8 dB) from an outdoor base to a terminal just inside a high-efficiency building is under 35m.

ACKNOWLEDGMENT

The authors would like to thank Mikko Uusitalo for the support of the measurement campaign. This project has received funding from the European Union's Framework Programme for Research and Innovation Horizon 2020 under Grant Agreement No. 861222 (EU H2020 MSCA MINTS). Rodolfo Feick acknowledges the support of ANID PIA/APOYO AFB180002.